\documentclass[twocolumn]{ceurart}

\usepackage{listings}
\usepackage{xcolor}
\usepackage{balance}

\begin{document}

\copyrightyear{2025}
\copyrightclause{Copyright for this paper by its authors. Use permitted under Creative Commons License Attribution 4.0 International (CC BY 4.0).}
\conference{DOLAP 2025: 27th International Workshop on Design, Optimization, Languages and Analytical Processing of Big Data, co-located with EDBT/ICDT 2025, March 25, 2025, Barcelona, Spain}

\title{The Case for Instance-Optimized LLMs in OLAP Databases}

\author[1]{Bardia Mohammadi}[%
orcid=0009-0001-3658-7291,
email=bmohamma@mpi-sws.org,
url=https://bardia-mhd.github.io/,
]
\cormark[1]
\author[1]{Laurent Bindschaedler}[%
orcid=0000-0003-0559-631X,
email=bindsch@mpi-sws.org,
url=https://binds.ch,
]
\address[1]{Max Planck Institute for Software Systems, Saarbrücken, Germany}

\cortext[1]{Corresponding author.}


\begin{abstract}
Large Language Models (LLMs) can enhance analytics systems with powerful data summarization, cleaning, and semantic transformation capabilities. However, deploying LLMs at scale -- processing millions to billions of rows -- remains prohibitively expensive in computation and memory. We present IOLM-DB, a novel system that makes LLM-enhanced database queries practical through query-specific model optimization. Instead of using general-purpose LLMs, IOLM-DB generates lightweight, specialized models tailored to each query's specific needs using representative data samples. IOLM-DB reduces model footprints by up to 76\% and increases throughput by up to 3.31$\times$ while maintaining accuracy through aggressive compression techniques, including quantization, sparsification, and structural pruning. We further show how our approach enables higher parallelism on existing hardware and seamlessly supports caching and batching strategies to reduce overheads. Our prototype demonstrates that leveraging LLM queries inside analytics systems is feasible at scale, opening new possibilities for future OLAP applications.
\end{abstract}


\maketitle

\section{Introduction}
LLMs have demonstrated exceptional capabilities in natural language understanding and generation~\cite{few-shot-learners,gemini,llama,sparks-agi}. One promising application in data management is integrating LLM prompting into database queries. This approach is particularly useful in analytical database systems, enabling users to harness the power of LLMs directly in their queries~\cite{data-wrangling,can-llm-wrangle}. For instance, given a table of unstructured product reviews, a user could write a query like:

\begin{lstlisting}[language=SQL]
SELECT product_id, user_id, 
       prompt('summarize in 5 words: ' || review) AS 
    review_summary 
FROM product_reviews;
\end{lstlisting}

\noindent This approach opens up new possibilities for generating, summarizing, cleaning, and transforming structured and unstructured data directly within the database.

However, applying LLM prompts to each row of data presents significant challenges. A typical query, such as a transformation, classification, or schema extraction, requires a separate LLM invocation involving tokenization, context encoding, and autoregressive decoding. This per-row inference results in high computational overhead, as even simple queries can trigger millions or billions of LLM calls, leading to excessive latency and resource consumption, especially for large tables. While running local models~\cite{llama,mistral} instead of cloud-based models~\cite{gpt4,claude,gemini} may reduce costs and latency, the overheads typically remain significant relative to conventional database operations, and running large-scale distributed models may prove challenging due to the lack of co-located accelerators and large memory requirements. These limitations highlight the importance of developing efficient strategies to integrate LLMs into analytics workflows without compromising performance or scalability.

We propose a practical approach to mitigate these challenges: instance-optimized LLMs for databases (IOLM-DB). By tailoring models to the specific workloads and data distributions of a given query and database instance, we reduce the cost of LLM inference, making it more practical to use at scale. We find that the OLAP setup is an ideal environment for creating optimized models because it operates in a controlled setting where the workload and data are predictable. IOLM-DB combines multiple model compression techniques, including quantization (reducing numerical precision)~\cite{egashira2024exploiting}, sparsification (introducing zero elements)~\cite{Frantar2023SparseGPTML}, and structural pruning (removing non-essential components)~\cite{Ma2023LLMPrunerOT}. The resulting models preserve task-relevant capabilities while being significantly smaller and cheaper to execute with higher parallelism, helping to narrow the performance gap for row-by-row LLM execution.

We have developed an initial prototype of IOLM-DB that targets Python's \texttt{pandas} library~\cite{pandas} for rapid iteration. By working in this simplified environment, we can experiment with various optimization strategies and gather initial performance insights before integrating these strategies into a production environment. The main objective of this paper is to provide a proof-of-concept for our approach, allowing us to evaluate potential performance gains and identify challenges that may arise when implementing and deploying instance-optimized models. Our preliminary results indicate that IOLM-DB can generate compressed models on the fly that are up to 3.28$\times$ smaller than the base model while sporting similar or better accuracy, achieves higher parallelism on the same hardware, and increases throughput between 2.52$\times$ and 3.31$\times$ on three representative workloads.

The paper makes the following contributions:
\vspace{-1mm}
\begin{itemize}
    \item We propose an end-to-end system for prototyping LLM prompting in OLAP scenarios and a series of workloads to assess its performance.
    \item We introduce the first method for generating instance-optimized LLMs in database environments.
    \item  We evaluate the efficiency of our method on the proposed workloads, showing significant performance improvements. These results indicate a promising first step toward demonstrating the feasibility of LLM compression for such applications.
\end{itemize}


\vspace{-2mm}
\paragraph{Artifact Availability} The source code and datasets are available at \url{https://github.com/mpi-dsg/IOLM-DB}.

\section{Background and Motivation}

This section briefly overviews key concepts and related work that motivate our approach.

\paragraph{LLM Prompting in Databases}
Augmenting databases with the ability to prompt LLMs can facilitate data wrangling and analysis by enabling natural language queries and transformations~\cite{can-llm-wrangle}. Rather than exporting data for external processing or implementing complex client-side integrations, bringing LLM capabilities directly into the database execution environment offers a more streamlined approach. Recent non-peer-reviewed work, such as LOTUS~\cite{patel2024lotus}, has demonstrated the feasibility of extending relational models with LLM-powered semantic operators, enabling AI-based operations like natural language-driven sorting and aggregation. This paper considers a system that integrates LLM prompts as first-class operations within the query processing pipeline, allowing them to be composed with traditional database operations. While our focus is on OLAP systems, the principles extend to other database architectures.
\vspace{-2mm}
\paragraph{New LLM-Based Capabilities}
LLM integration enables powerful new capabilities for database systems, especially when working with unstructured or semi-structured data. These capabilities include summarization, sentiment analysis, data extraction, error correction, and semantic transformations. LLMs can also enable more flexible abnd intuitive operations, such as fuzzy matching and semantic joins beyond exact string matching.
\vspace{-2mm}
\paragraph{Other Approaches} 
Existing approaches to adding these capabilities have notable limitations. Code generation techniques~\cite{evaluating-llm-code,data-wrangling}, where LLMs generate executable database code, can be brittle and struggle with complex transformations that require deep semantic understanding. Alternative approaches using simpler models trained on input-output pairs~\cite{flashfill++,data-extraction-regex,tde} face challenges handling diverse scenarios and require extensive training data curation. While these methods can work for specific use cases, they often fail to provide the flexibility and generality needed for broad adoption in database systems.
\vspace{-2mm}
\paragraph{The Need for Instance-Optimization}
We argue that anything short of directly invoking the LLM for such tasks is inherently limiting, as it would restrict the system's expressiveness. Therefore, efficiency is paramount to making LLMs practical at scale, especially for large-scale analytics. Many anticipated use cases require invoking the LLM at a fine-grained level, such as once per row or more frequently~\cite{can-llm-wrangle}. This granularity requires a new approach that reduces inference costs while maintaining accuracy, which motivates our approach of specializing LLMs for specific prompts and data patterns.

OLAP environments are particularly well-suited for such optimizations, as their controlled setting often allows query patterns and data characteristics to be inferred or extracted in advance, enabling model optimization tailored to these patterns. Similarly, interactive queries frequently exhibit recurring or predictable patterns that can be leveraged to create instance-optimized models, improving efficiency and accuracy as these patterns evolve over time. Although our current method may introduce too much overhead for some use cases, we expect it to be highly effective for long-running queries, where significant performance gains outweigh the upfront cost of optimization during execution.
\vspace{-2mm}
\paragraph{LLM Compression Techniques}
Recent advances in LLM optimization provide a foundation for our approach. We leverage three key techniques from the literature: quantization, which reduces numerical precision to decrease memory requirements~\cite{Frantar2022GPTQAP}; sparsification, which introduces strategic zero elements to minimize computational overhead~\cite{Frantar2023SparseGPTML}; and structural pruning, which removes non-essential model components~\cite{Ma2023LLMPrunerOT}. While these techniques have proven effective for domain-specific optimization, our scenario presents unique challenges. We need to optimize at a much finer granularity -- for individual queries or prompts rather than broad domains -- and ensure consistent, predictable behavior when working with structured data.

\vspace{-2mm}
\section{System Design and Architecture}

IOLM-DB is an OLAP system that integrates LLM invocation directly into its execution pipeline while ensuring that this integration is efficient, scalable, and cost-effective.
\vspace{-2mm}
\subsection{Overview}

We assume a setup where the user executes queries over one or more database tables (in our prototype, \texttt{pandas} DataFrames). These queries can include calls to an LLM, for example, to summarize free-form text columns, transform semi-structured fields, or provide semantic annotations. Rather than invoking a general-purpose LLM repeatedly and incurring high computational and memory overhead at runtime, IOLM-DB creates a specialized, instance-optimized LLM explicitly tailored for the query and dataset. For example, a query performing text summarization on product reviews requires a different optimization than a data correction query. To achieve this specialization, the model is compressed and pruned based on the query type and data characteristics. This approach is necessary because a one-size-fits-all model is inefficient; different queries demand varying levels of understanding, precision, and computational cost.

IOLM-DB works in tandem with the underlying analytics engine. While the engine handles standard relational operations efficiently, we introduce custom operators that intercept LLM prompts in the query. These operators trigger a workflow that generates a specialized LLM for that particular query and data distribution. This process leverages a suite of techniques -- quantification, sparsification, and pruning -- to minimize both the memory footprint of the LLM and the inference costs, ensuring that at runtime the optimized LLM can be invoked with minimal latency and resource usage.

We drastically reduce per-row inference overhead by producing a specialized LLM for each query instance. This approach enables running LLM-based transformations and analyses at scale, supporting massive parallelism and efficient resource utilization. Ultimately, the goal is to handle large workloads and large numbers of concurrent queries while providing near-interactive response times.
\vspace{-2mm}
\subsection{Techniques to Generate Instance-Optimized Model}

Our system combines multiple compression techniques to reduce memory consumption and compute costs at the LLM level. By carefully integrating these methods, we maintain model accuracy while achieving significant resource savings. We combine the following techniques in IOLM-DB.
\paragraph{Quantization} By lowering numerical precision (e.g., using 8-bit weights and sometimes 8-bit activations), we reduce both the GPU memory footprint and computational overhead. Quantization retains the model's core capabilities but significantly increases inference throughput.
\vspace{-2mm}
\paragraph{Sparsification} Imposing structured or unstructured sparsity patterns in model weights reduces the number of active parameters, leading to fewer operations during inference and, in some cases, allows for hardware-level support for sparse computations.
\vspace{-2mm}
\paragraph{Structural Pruning} Removing entire components such as attention heads or entire layers that contribute little to the task at hand reduces model depth and complexity. Full structural pruning, aided by tools such as LLM-Pruner~\cite{ma2023llm}, enables the construction of ultra-compact models specialized for the given query patterns.
\noindent
These techniques are not applied in isolation. Methods such as GPTQ~\cite{frantar2023gptq} and SmoothQuant~\cite{xiao2023smoothquant} combine pruning and quantization steps to preserve accuracy while aggressively reducing size. SparseGPT~\cite{Frantar2023SparseGPTML} applies pruning strategies suitable for large models in a one-shot manner, maintaining accuracy even at high sparsity levels.
\noindent
The result is a highly compressed model tailored to the query's distribution and the dataset's characteristics. We use calibration data -- small, unlabeled samples representing the query's input domain -- to fine-tune quantization parameters and pruning thresholds. This process ensures that the specialized LLM efficiently handles the target data, reducing resource requirements while minimizing accuracy loss.
\vspace{-2mm}
\subsection{Runtime and Invocation Optimizations}

At runtime, the optimized LLM is invoked by custom operators within the OLAP engine's execution pipeline. IOLM-DB uses the following optimizations to further reduce overhead.
\vspace{-2mm}
\paragraph{Caching} Intermediate results and repeated inputs are cached so that identical LLM queries need not be recomputed, which is especially valuable when data contains frequent duplicates or recurring patterns.
\vspace{-2mm}
\paragraph{Batching} We batch multiple requests to the LLM together to amortize invocation overhead. By grouping multiple rows or operations, we minimize switching costs and achieve higher throughput.
\vspace{-2mm}
\paragraph{Cascading (Future Work)} We plan to explore cascading strategies in the future, where an initial coarse-grained LLM invocation feeds into more specialized or higher-precision models only where needed. This approach could further refine trade-offs between speed, cost, and accuracy.

\vspace{-3mm}
\section{Workloads and Use Cases}
\label{sec:workloads}

We evaluate our approach on three representative workloads, each designed to highlight a different aspect of integrating LLM invocations into OLAP queries. All these workloads operate on real-world datasets (e.g., Amazon Review~\cite{Ni2019JustifyingRU}, Github Typos~\cite{hagiwara-mita-2020-github}) and are designed to stress different points along the performance-accuracy spectrum.
\vspace{-3mm}
\subsection{Summarization (Text Reduction)} This workload involves condensing verbose free-text fields into short summaries. An example use case is taking product reviews, extracting their essential meaning, and outputting a concise summary -- e.g., summarizing each product review into five words. Such summarization helps analysts quickly gain insights from large volumes of textual data without manually sifting through lengthy entries. By operating on unstructured Amazon Reviews datasets~\cite{Ni2019JustifyingRU}, we test the system's ability to scale and maintain accuracy when transforming large amounts of unstructured text.
\vspace{-2mm}
\subsection{Data Correction} Data correction enhances data quality by addressing errors or inconsistencies. For instance, we can provide a specific data type or category and task the system with correcting typos or mismatches. By applying a per-row invocation of the language model, we can correct misspelled code or text records in GitHub for subsequent analysis~\cite{hagiwara-mita-2020-github}.
\vspace{-2mm}
\subsection{Fuzzy Joins (Semantic Mapping)} Fuzzy joins address the problem of integrating data from multiple tables by understanding semantic similarity rather than relying on exact string matches. For example, the system can determine whether two entries from different datasets refer to the same entity despite slight differences in wording or formatting, even when textual fields do not perfectly align, improving data integration and discovery.

\vspace{-2mm}
\section{Evaluation}







We conduct a preliminary evaluation of IOLM-DB through experiments on the three workloads and associated datasets described in Section~\ref{sec:workloads}. Our key objective is to verify the viability of our approach for scaling LLM invocations per row in a realistic setting.
\vspace{-2mm}
\paragraph{Metrics}

Our evaluation centers on the following metrics:
\begin{itemize}
    \item \textbf{Throughput:} The number of rows processed per second by the system, which reflects its overall efficiency and scalability.
    
    \item \textbf{Model Size:} The size of the model, which serves as a proxy for GPU memory usage and its capacity to parallelize execution effectively. Smaller models typically reduce memory pressure and improve resource utilization.
    
    \item \textbf{Accuracy:} The proportion of rows where the system produces correct results. For this evaluation, we assume the baseline model achieves perfect accuracy (accuracy = 1), and we compare the optimized models by normalizing against this standard.
\end{itemize}
\vspace{-3mm}
\paragraph{Models}

We conduct our evaluation using Meta's Llama 3.1 Instruct model with 8 billion parameters instruction-tuned language model (Llama-3.1-Instruct-8B)~\cite{llama}, which strikes a balance between size and performance for the workloads under consideration. The model is compact enough to fit on a single modern GPU yet powerful enough to deliver strong performance across all tasks. We verified our Llama model's suitability for these tasks by comparing it against OpenAI's GPT4o and Anthropic's Claude Sonnet 3.5 and manually verifying the results.

\paragraph{Baseline and Configurations}
We execute IOLM-DB using Llama-3.1-Instruct-8B, which serves as our baseline for all metrics. Building upon this foundation, we evaluate two optimized variants:
\begin{itemize}
    \item IOLM-DB-Perf: instance-optimized version of the Llama-3.1-Instruct-8B with the best throughput.
    \item IOLM-DB-Acc: instance-optimized version of the Llama-3.1-Instruct-8B with the best accuracy. 
\end{itemize}

These two variants highlight an interesting trade-off between performance and accuracy enabled by IOLM-DB which we aim to explore in future work, allowing the user to select whether to spend more time for better results or to produce slightly less accurate results faster.
\vspace{-2mm}
\paragraph{Hardware and Configuration}

We run these experiments on a machine equipped with an NVIDIA Hopper H100 80~GB SXM GPU, two AMD CPUs (EPYC 9654), and 2~TB of DDR5. The operating system is 64-bit Debian Linux (kernel version 5.15). We use CUDA 11.8, vLLM 0.6.3, and HuggingFace 4.46.1 transformers.

\subsection{Performance, Model Size, and Accuracy}

Table~\ref{tab:model_comparison} compares the baseline with our IOLM-DB instance-optimized models. We show the overall model size as measured on the GPU, the resulting accuracy in our dataset (normalized to the baseline), and the throughput achieved in each case (in rows processed per second).

\begin{table*}
\begin{tabular}{|l|l|c|c|c|}
\hline
\textbf{Workload} & \textbf{Model} & \textbf{Model Size} & \textbf{Accuracy Score} & \textbf{Throughput} \\
\hline
\multirow{3}{*}{Summarization} & Baseline & 14.98 GB & 1 & 4.67 rows/s \\
& IOLM-DB-Perf & 8.48 GB & 0.91 & 15.50 rows/s \\
& IOLM-DB-Acc & 8.48 GB & 1 & 11.97 rows/s \\
\hline
\multirow{3}{*}{Data Correction} & Baseline & 14.98 GB & 1 & 2.73 rows/s \\
& IOLM-DB-Perf & 8.48 GB & 1 & 7.60 rows/s \\
& IOLM-DB-Acc & 8.48 GB & 1 & 7.60 rows/s \\
\hline
\multirow{3}{*}{Fuzzy Join} & Baseline & 14.98 GB & 1 & 14.92 rows/s \\
& IOLM-DB-Perf & 8.48 GB & 1 & 37.72 rows/s \\
& IOLM-DB-Acc & 8.48 GB & 1 & 37.72 rows/s \\
\hline
\end{tabular}
\caption{Throughput, memory usage, and accuracy score (normalized to the baseline) for each of the three workloads considered in this paper. IOLM-DB-Perf corresponds to the compressed model with the highest throughput, while IOLM-DB-Acc is the compressed model with the highest accuracy score. We report the overall throughput in rows per second.}
\label{tab:model_comparison}
\end{table*}

These results underscore the benefits of instance-optimization for LLMs across all workloads considered. In all cases except summarization, both IOLM-DB-Perf and IOLM-DB-Acc achieve an accuracy score of 1, demonstrating that our approach maintains baseline-level accuracy. This outcome highlights that our optimizations preserve model quality while reducing resource requirements.

Our method achieves significant reductions in model size, with compression factors of up to 76\%, through quantization, sparsification, and pruning. These reductions lower memory requirements and improve computational efficiency. The throughput improvements are substantial: 2.78$\times$ for data correction and 2.52$\times$ for fuzzy join workloads. For summarization, despite a slight accuracy decrease in IOLM-DB-Perf, we observe a 3.31$\times$ throughput improvement. These performance gains stem from the compressed models and reduced model size, enabling higher parallelism.

The results demonstrate that instance-optimized LLMs can effectively balance accuracy and efficiency, making them suitable for performance-critical applications.
\vspace{-2mm}
\subsection{Discussion}

Although these results are preliminary, they highlight the potential of instance-optimization to enable efficient LLM querying in analytics workloads. Our evaluation also revealed several areas for improvement that could further enhance the performance and scalability of our solution.

One key bottleneck identified in our initial prototype is the space consumed by the vLLM cache. Optimizing the cache design could further reduce memory usage and enable increased parallelism, particularly for workloads with high concurrency demands. Also, the interface between the \texttt{pandas} library and our instance-optimized operators presents another area for improvement. Reengineering this interface to reduce overhead and improve integration could significantly boost overall performance. We expect to gain one or two orders of magnitude performance improvements from these optimizations.

An interesting observation is that the resulting quality, particularly for IOLM-DB-Acc, often improves compared to the baseline. However, the normalization of the accuracy score prevents this improvement from being reflected in Table~\ref{tab:model_comparison}. These findings suggest that our instance-optimized approach not only preserves accuracy but can, in some cases, enhance it. Further investigation is needed to understand the underlying factors driving these improvements and to explore how they can be consistently leveraged.

Finally, we plan to explore additional techniques in the LLM compression space, such as advanced quantization methods and knowledge distillation. The time to generate an instance-optimized LLM is in the order of single-digit minutes, which we believe is acceptable for most table sizes. However, reducing that further could unlock additional use cases outside the OLAP space.
\vspace{-2mm}
\section{Conclusion}
\vspace{-2mm}
This paper presented IOLM-DB, a proof-of-concept system for integrating LLM operations directly into OLAP query pipelines. The core innovation lies in instance-optimizing LLMs based on the specific query and dataset at hand. IOLM-DB can generate specialized models that drastically reduce inference costs without sacrificing accuracy by leveraging quantization, sparsification, and pruning techniques. Our preliminary results suggest that this approach makes row-by-row invocation of LLMs at scale both practical and efficient, bridging the gap between powerful linguistic transformations and large-scale analytics.

This line of work opens up promising research directions, encouraging the community to refine these techniques further, explore new instance-level optimizations, and ultimately bring high-performance, query-aware LLM capabilities into mainstream analytical workflows.

\bibliography{main}

\end{document}